\documentclass[slac_one]{revtex4}
\usepackage{graphicx}
\usepackage{fancyhdr}
\begin{document}

\title{GammeV: a gamma to milli-eV particle search
at Fermilab} 

%

\author{W. Wester}
\affiliation{Fermilab, Batavia, IL 60510, USA \\
presented on behalf of the GammeV Collaboration}

\begin{abstract}
GammeV is an experiment conducted at 
Fermilab that employs the light shining through a wall technique to
search for axion-like particles and employs a particle in a jar
technique to search for dilaton-like chameleon particles. We obtain
limits on the coupling of photons to an axion-like particle that extend
previous limits for both scalars and pseudoscalars in the 
milli-eV mass range. We are able to exclude the axion-like
particle interpretation of the anomalous PVLAS 2006 result by more than 5
standard deviations. We also present results on a search for chameleons and
set limits on their possible coupling to photons.
\end{abstract}

\maketitle

\thispagestyle{fancy}


\section{Introduction}

{\em What are the particles that make up the dark matter
of the universe?}, is one of the most fundamental scientific questions today.
Axion-like 
particles or other weakly interacting sub-eV particles (WISPs)
are highly motivated dark matter particle candidates
since they have properties that might explain the cosmic abundance of dark matter.
The milli-eV mass scale arises in several areas of modern particle physics
with a see-saw between the Planck and TeV
mass scales, neutrino mass differences, the dark energy density expressed 
in milli-eV$^{4}$, and known dark matter candidates. 
In 2006, the PVLAS experiment reported \cite{pvlas2006} (although no
longer observes \cite{pvlas2007})
anomalous polarization effects on an incident laser
in the presence of a magentic field which could be
interpretated as being due to an axion-like particle
in the milli-eV mass range with a unexpectedly strong coupling to photons.

A previous laser experiment (BFRT) in the early 1990's
used a ``light shining through a wall'' (LSW) \cite{lsw}
technique to set limits on sub-eV axion-like
particles \cite{bfrt}.
Because this experiment used available 4.4m long magnets, they had regions
with no sensitivity for an axion-like particle in the mass
range suggested by the anomalous PVLAS result. The GammeV experiment \cite{gammevweb}
has been proposed
to examine the milli-eV mass scale for an axion-like particle that couples to
photons to resolve the possible mystery of the anomalous PVLAS result and to extend
the search for axion-like and chameleon particles 
in the milli-eV mass region.

\section{GammeV Apparatus}

\begin{figure}[hb]
\begin{minipage}{0.45\linewidth}
\centerline{\includegraphics[width=\textwidth]{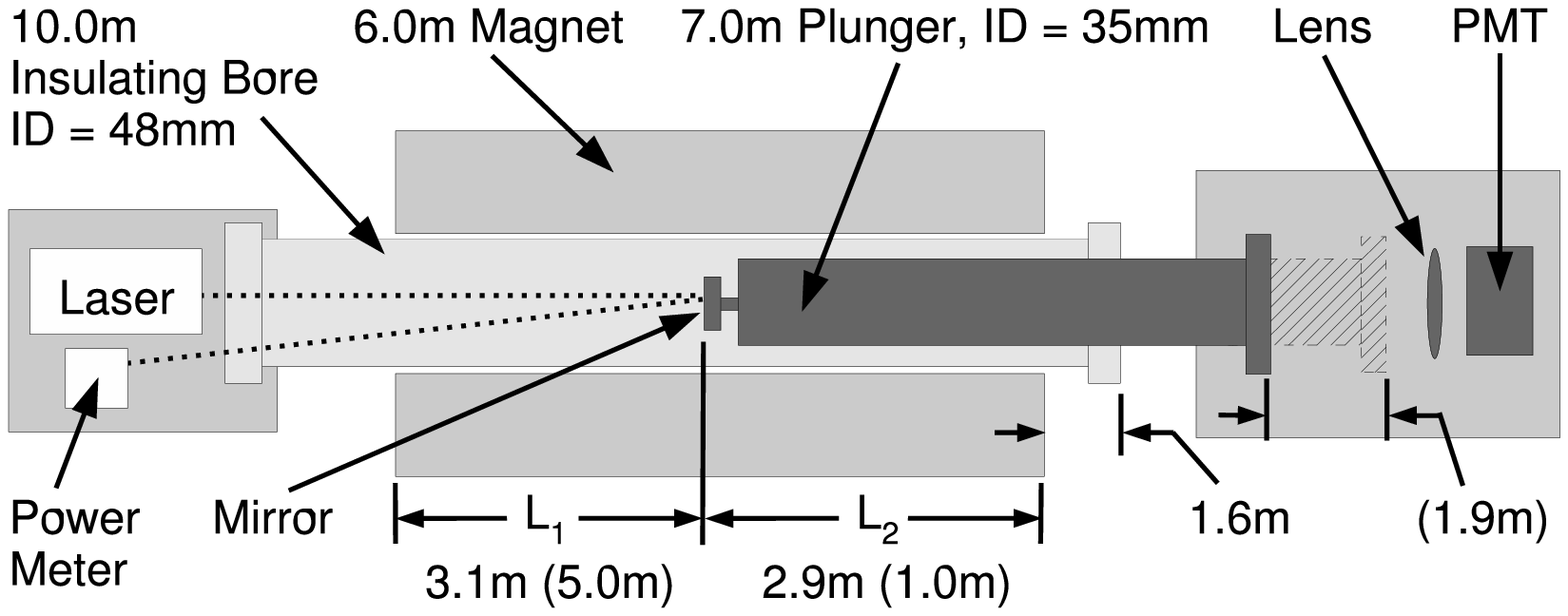}}
\caption{Schematic diagram of the GammeV experimental apparatus showing a 
frequency-doubled Nd:YAG
laser sending 20Hz of 10~ns wide laser pulses down the warm bore of a Tevatron
dipole magnet. In either the middle of the magnet or towards one end is the ``wall''
which reflects the laser back onto a power meter. The ``wall'' is mounted on
a sliding vacuum tube, the plunger, which is welded light-tight inside the magnet and
which extends into a PMT dark box. For the particle in a jar chameleon search,
the wall is removed and the mirror is placed in the PMT box. The laser is
then turned off, the mirror removed, and the PMT is turned on to search for 
an afterglow of possible regenerated photons.}
\label{Fig:apparatus}
\end{minipage}
\hspace{0.2cm}
\begin{minipage}{0.45\linewidth}
\centerline{\includegraphics[width=\textwidth]{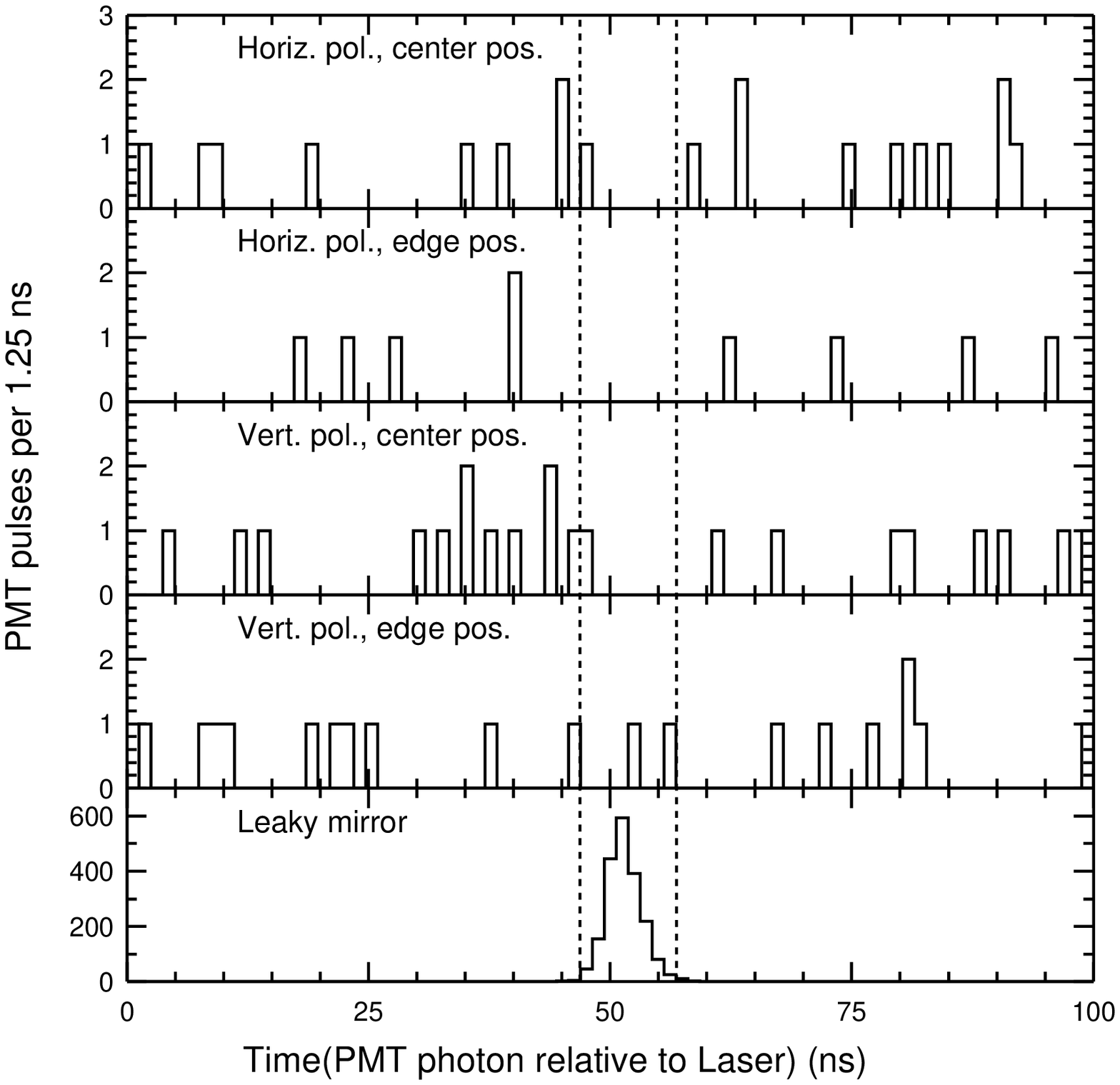}}
\caption{PMT pulse time relative to laser pulse time for the four run configurations
shown relative to the expected time distribution of photons form the ``leaky
mirror'' calibration data.}
\label{Fig:data}
\end{minipage}
\end{figure}

The GammeV LSW apparatus is shown schematically in Fig.~\ref{Fig:apparatus} 
where, in the presence of an external magnetic field,
a laser photon might oscillate into an
axion-like particle that can traverse a ``wall'' and then have a small 
probability to regenerate back into a detectable photon. The formula for
the probability of this regeneration is given by the following:
\begin{eqnarray}
\label{E:regenprob1}
P_{regen} &=&  \frac{16 B_1^2 B_2^2\omega^4}{M^4 m_\phi^8} \sin^2 \left( \frac{m_\phi^2 L_1}{4\omega} \right) \cdot \sin^2 \left( \frac{m_\phi^2 L_2}{4\omega} \right) \nonumber \\
\label{E:regenprob2}
&=& (2.25\times 10^{-22}) \times \frac{(B_1/\mbox{Tesla})^2(B_2/\mbox{Tesla})^2(\omega/\mbox{eV})^4}{(M/10^5 \mbox{ GeV})^4 (m_\phi/10^{-3} \mbox{ eV})^8} \nonumber\\
 & & \times \sin^2 \left( 1.267 \frac{(m_\phi/10^{-3} \mbox{ eV})^2 (L_1/\mbox{m})}{(\omega/\mbox{eV})} \right) \sin^2 \left( 1.267 \frac{(m_\phi/10^{-3} \mbox{ eV})^2 (L_2/\mbox{m})}{(\omega/\mbox{eV} \nonumber)} \right)
\end{eqnarray}
where $\omega$ is the photon energy, $M$ is a high mass scale inverse to the
coupling to photons $g_{a\gamma\gamma}$, $m_{\phi}$ is the mass of the axion-like 
particle, and $B_1$, $L_1$, $B_2$ and $L_2$ are the magnetic field strengths and 
lengths in the photon conversion and regeneration regions, respectively.

The GammeV experiment utilizes two novel aspects in order to have increased
sensitivity over the region of interest. The plunger is constructed so that it
can place the ``wall'' either in the middle ($L_1 = L_2$)
of the magnet or towards one end
of the magnet ($L_1 \neq L_2$). Regions of 
insensitivity will be shifted when
the plunger is put in two distinct positions and the entire milli-eV 
mass range can be probed with high sensitivity. The second aspect
is to utilize time correlated single photon counting techniques 
where regenerated
photons are searched in a 10~ns wide window that is intrinsically low
noise since the appropriate overlap of a laser pulses with dark pulses
from the PMT is low over the duration of the experiment.

\section{Data Taking and Axion-like Particle Results}

GammeV has recorded calibration data called the ``leaky mirror'' sample 
where the 
wall was removed and the each laser pulse was attenuated by $\sim 10^{19}$ 
so that single photons could be recorded
by the PMT. The end result was that we confirm
the speed of light travel of those photons through the GammeV apparatus, 
and establish a 10~ns wide window
{\it a priori} for our search region since regenerated photons from an axion-like
particle would have the same relative timing. By putting a polarizing filter
before the PMT, we also verified that the polarization of the laser light was parallel
or perpendicular to the magnetic field depending on whether we inserted a 1/2-wave
plate into the optical path. We would thus be able to probe either scalar or
pseudoscalar axion-like particles which differ in that the coupling requires
the polarization to be perpendicular to, or aligned with the magnetic field.

Data was acquired in four configurations: two polarizations and two positions of the 
``wall.'' In each configuration, approximately 20 hours of data was acquired - nearly
1.5M pulses with 4$~\times~10^{17}$ photons per pulse. The time of each laser pulse
was recorded along with the time of each pulse detected by the PMT. In an offline
analysis of the data, the timing of PMT pulses relative to the laser pulse could be
examined in the temporal region where the ``leaky mirror'' calibration photons
were also recorded. 
Fig.~\ref{Fig:data} shows the data recorded in the four configurations
where 1, 0, 1, and 2 signal candidates are observed in the 10~ns wide search window.
The expected background is obtained from the data by looking at the number of
PMT pulses within 10000~ns of the laser firing and indicates that we should have
expected approximately 1.5 events of expected background in each of the four
configurations. The data show no indication of
regenerated photons above this background. We use the determined efficiencies of the PMT and optical
transport along with the measured laser power for each pulse to obtain the normalization
on the number of incident photons and the expectation of the signal rate. We
account for systematic uncertainties in these quantities in the derrived
limits.

\begin{figure}
\begin{minipage}[hb]{0.45\linewidth}
\centerline{\includegraphics[width=\textwidth]{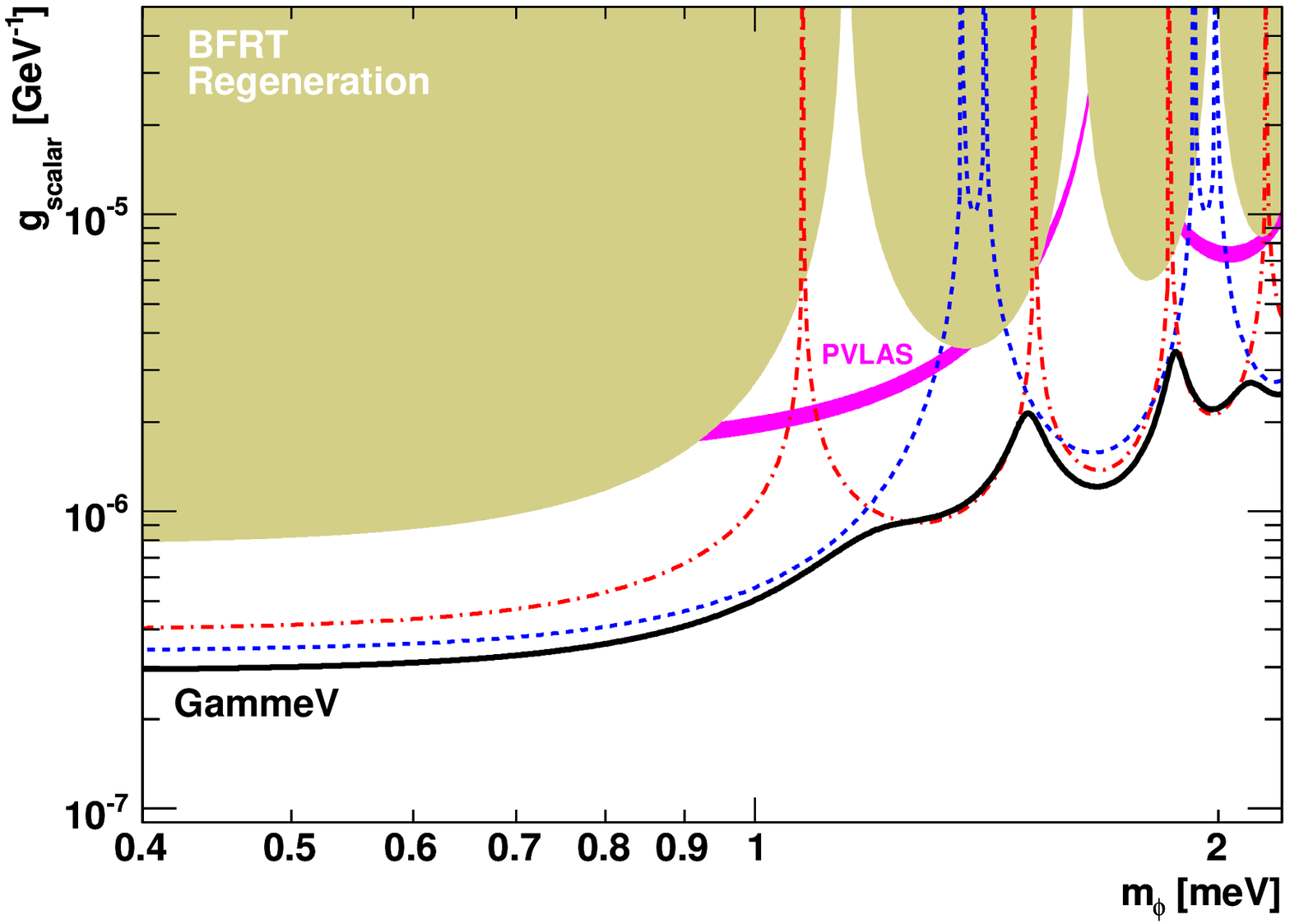}}
\caption{Exclusion region obtained by the GammeV
for the coupling to photons versus the mass of a scalar axion-like
particle. The dashed lines show the limits obtained separately for the data
recorded with the ``wall'' in the middle and near one end of the maget. Also
shown is the anomalous PVLAS region of interest. Finally,the shaded region indicates 
the previous
exclusion from BFRT.}
\label{Fig:scalar}
\end{minipage}
\hspace{0.2cm}
\begin{minipage}[hb]{0.45\linewidth}
\centerline{\includegraphics[width=\textwidth]{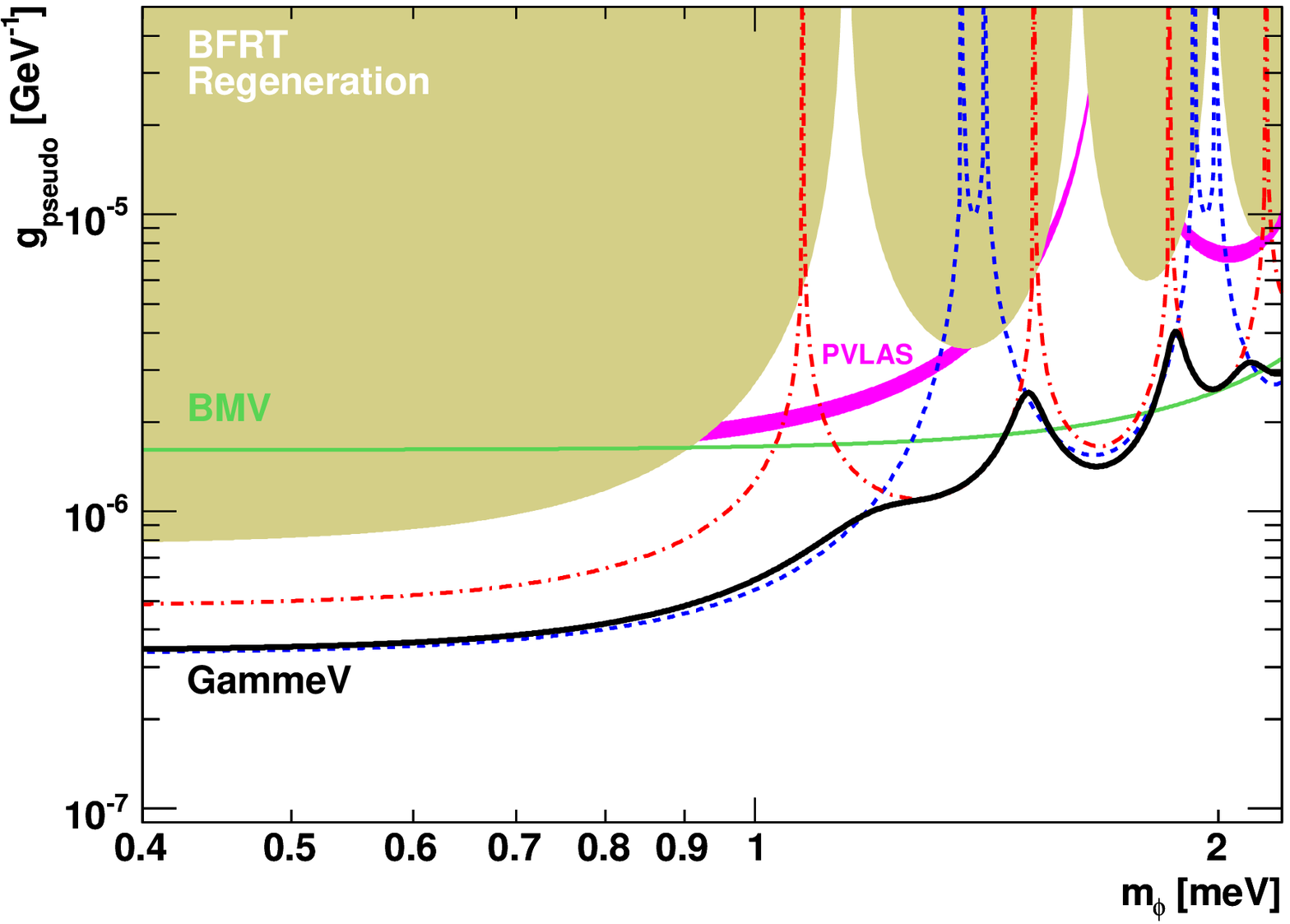}}
\caption{Exclusion region for the case that the axion-like particle couples
to photons as a pseudoscalar. Also shown is a recent limit obtained by the
BMV \cite{bmv} experiment.}
\label{Fig:pseudo}
\end{minipage}
\end{figure}

The 
non-observation of an excess signal allows us to set limits at 3$\sigma$ of the 
axion-like particle coupling to photons versus the mass that extend the 
previously excluded
region. In addition, we exclude at more than 5$\sigma$ the axion-like
particle interpretation of the PVLAS anomoly. Figures~\ref{Fig:scalar},~\ref{Fig:pseudo} 
show the
resulting limits for the coupling of scalar and pseudoscalar axion-like particles to
photons in milli-eV mass region. Other recent experiments that also probe
this region of interest have also reported null results \cite{OtherExpts}.

\begin{figure}[hb]
\centerline{\includegraphics[width=0.4\textwidth,angle=-90]{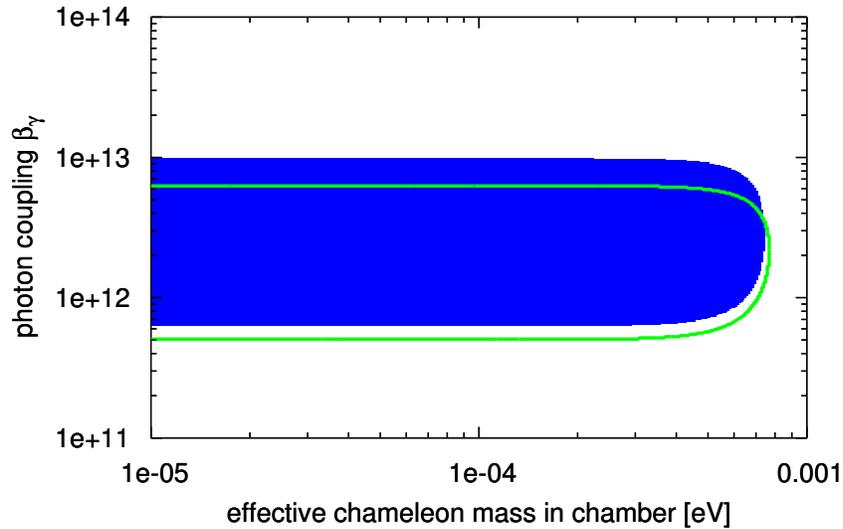}}
\caption{Chameleon exclusion region obtained by the GammeV
for the coupling to photons (expressed relative to the reduced Planck mass) 
versus the effective mass of the chameleon in the vacuum of the 
GammeV apparatus.}
\label{Fig:ChamLimits}
\end{figure}

\section{Chameleons}

Chameleons are possible fundamental (pseudo-)scalars that might explain the
dark energy of the unvierse with the property that they couple to the
stress energy tensor \cite{ChamTheory} which 
allows them to change their properties 
depending on their environment. If chameleons couple to photons they could
be generated by a laser shining through a magnetic field and be 100\% 
reflective upon encountering ordinary matter. Thus they could be trapped
in a jar as they bounce between the windows and vacuum tubes
of the GammeV apparatus. After
turning off the laser and removing the wall, the PMT can be turned on to
see whether the trapped chameleons might produce a detectable afterglow
by having a rate for regenerating back into photons. GammeV searched for
such an afterglow and set limits under various assumptions of the form
of the chameleon potential on the possible coupling to photons versus
an effective chameleon mass \cite{GammeVCham}. These limits are
shown in Fig.~\ref{Fig:ChamLimits}. More
details on the region of validity for these limits can be found in 
\cite{GammeVCham}.

\section{Future Prospects}

There is continued motivation to search for possible sub-eV (pseudo-)scalars
that couple to photons as a possible contribution to the dark matter of
the universe. A next generation chameleon experiement will use a modified
vacuum system. For ``light shining through a wall'' experiments, one
would like to improve limits by more than three orders of magnitude to
probe a region previously unexplored by laboratory or astrophysical data.
This might be possible by utilizing two phase-matched Fabry-Perot optical
cavities on both the generation and regeneration sides of the wall
in order to resonnantly enhance the axion-like particle to photon
regeneration. \cite{EnhancedGammeV}.

\begin{acknowledgments}
This work is supported by the U.S. Department of Energy under Contract No.
DE-AC02-07CH11359. The project is made possible through the work of technical
resources including a mechanical team within the Fermilab Particle Physics
Division and a magnet team within the Fermilab Technical Division.
\end{acknowledgments}

\end{document}